%% file: main.tex
\documentclass[camera,letterpaper,nomarginnotes,nonarrowgutter]{jpaper}

\input{macros}

\begin{document}
%\urlstyle{tt}

\bstctlcite{IEEEexample:BSTcontrol}

\title{SimplePIM: A Software Framework {for} \\ Productive {and} Efficient {Processing-in-Memory}}

\author{Jinfan Chen$^1$\hspace{1em} Juan Gómez-Luna$^1$\hspace{1em} Izzat El Hajj$^2$ \hspace{1em} Yuxin Guo$^1$ \hspace{1em} Onur Mutlu$^1$ \\
\normalsize{
    $^1$ETH Zürich \hspace{1em} $^2$American University of Beirut
}}

%\date{}
\maketitle

%\thispagestyle{firstpage}
%\thispagestyle{empty}

%\IEEEpeerreviewmaketitle

\input{secs/0-Abstract}

%\keywords{processing-in-memory}

\input{secs/1-Introduction}

\input{secs/2-Background}

\input{secs/3-SmallTable.tex}
\input{secs/4-Implementation}
\input{secs/5-Experiments}

\input{secs/6-Discussion}

\input{secs/7-RelatedWork}

\input{secs/8-Conclusion}

\section*{Acknowledgments}
{We acknowledge support from the SAFARI Research Group’s industrial partners, especially Google, Huawei, Intel, Microsoft, VMware, and the Semiconductor Research Corporation. 
This research was partially supported by the ETH Future Computing Laboratory and the European Union's Horizon programme for research and innovation under grant agreement No. 101047160, project BioPIM (Processing-in-memory architectures and programming libraries for bioinformatics algorithms). 
This research was also partially supported by ACCESS – AI Chip Center for Emerging Smart Systems, sponsored by InnoHK funding, Hong Kong SAR.}

\bibliographystyle{IEEEtran}

\balance
\bibliography{refs}

\end{document}

%% file: macros.tex
\usepackage{amsmath}

\usepackage{hyperref}
\usepackage{balance}
\usepackage{caption}
\usepackage{subcaption}
\usepackage{graphicx}
\usepackage{textcomp}
\usepackage{subcaption}
\usepackage{multirow}
\usepackage[T1]{fontenc}
\usepackage{enumitem}
\usepackage{comment}
\usepackage{multirow}

\usepackage{algorithm,algorithmicx,algpseudocode}
\usepackage{listings}
\usepackage{pifont}
\usepackage{booktabs}
\usepackage{enumitem}
\setlist[itemize]{leftmargin=*}
\setlist[enumerate]{leftmargin=*}
\setlist{nolistsep}
\def\BibTeX{{\rm B\kern-.05em{\sc i\kern-.025em b}\kern-.08em
    T\kern-.1667em\lower.7ex\hbox{E}\kern-.125emX}}

\graphicspath{ {images/} }

\usepackage{xcolor}

\definecolor{codegreen}{rgb}{0,0.6,0}
\definecolor{codegray}{rgb}{0.5,0.5,0.5}
\definecolor{codepurple}{rgb}{0.58,0,0.82}
\definecolor{backcolour}{rgb}{0.95,0.95,0.92}

\lstdefinestyle{mystyle}{
  backgroundcolor=\color{backcolour}, commentstyle=\color{codegreen},
  keywordstyle=\color{magenta},
  numberstyle=\tiny\color{codegray},
  stringstyle=\color{codepurple},
  basicstyle=\ttfamily\footnotesize,
  breakatwhitespace=false,         
  breaklines=true,                 
  captionpos=b,                    
  keepspaces=true,                 
  numbers=left,                    
  numbersep=5pt,                  
  showspaces=false,                
  showstringspaces=false,
  showtabs=false,                  
  tabsize=2
}

\definecolor{mGreen}{rgb}{0,0.6,0}
\definecolor{mGray}{rgb}{0.5,0.5,0.5}
\definecolor{mPurple}{rgb}{0.58,0,0.82}
\definecolor{backgroundColour}{rgb}{0.95,0.95,0.92}

\lstdefinestyle{CStyle}{
    commentstyle=\color{mGreen},
    backgroundcolor=\color{backgroundColour},
    keywordstyle=\color{blue}\ttfamily,
    stringstyle=\color{red}\ttfamily,
    language=C,
    breaklines=true, 
    basicstyle=\ttfamily\footnotesize, 
}

\usepackage{tikz}

\definecolor{dkred}{rgb}{0.6,0.0,0}
\definecolor{dkgreen}{rgb}{0,0.5,0}
\definecolor{ddkgreen}{rgb}{0,0.7,0}
\definecolor{ddkblue}{rgb}{0,0,0.8}
\definecolor{gray}{rgb}{0.5,0.5,0.5}
\definecolor{mauve}{rgb}{0.58,0,0.82}

\newcommand{\jfc}[1]{\textcolor{black}{#1}}

\newcommand\juang[1]{\noindent{\color{black} {#1}}}

\newcommand\ignore[1]{ }

\AtBeginDocument{%
  \providecommand\BibTeX{{%
    \normalfont B\kern-0.5em{\scshape i\kern-0.25em b}\kern-0.8em\TeX}}}

\usepackage{cite}
\ignore{

\usepackage{fancyhdr}
\usepackage[en-GB, useregional=numeric]{datetime2}
\newcommand{\versionnum}[0]{2.1}
\pagestyle{fancy}
\fancyhf{}

\fancyhead[C]{\textcolor{ddkblue}{\emph{Version \versionnum~---\DTMnow}}}
\footskip=30pt
\fancyfoot{}
\fancyfoot[C]{\thepage}
    
\fancypagestyle{firstpage}
{
\fancyfoot[C]{\thepage}
}

}

%% file: secs/0-Abstract.tex
\begin{abstract}

Data movement between memory and processors is a major bottleneck in modern computing systems. 
The processing-in-memory (PIM) paradigm aims to alleviate this bottleneck by performing computation inside memory chips.
Real PIM hardware (e.g., the UPMEM system) is now available and has demonstrated potential in many applications.
However, programming such real PIM hardware remains a challenge for many programmers. 

This paper presents a new software framework, SimplePIM, to aid programming real PIM systems.
The framework processes arrays of arbitrary elements on a PIM device by calling iterator functions from the host and provides primitives for communication among PIM cores and between PIM and the host system.
We implement SimplePIM for the UPMEM PIM system and evaluate it on six major applications.
Our results show that SimplePIM enables 66.5\% to 83.1\% reduction in lines of code in PIM programs.
The resulting code leads to higher performance {(between 10\% and 37\% speedup)} than hand-optimized code in three applications and provides comparable performance in three others. 
{SimplePIM is fully and freely available at \url{https://github.com/CMU-SAFARI/SimplePIM}.} %\om{Repo ready and vetted?}

\end{abstract}

%% file: secs/1-Introduction.tex
%\def\thefootnote{*}\footnotetext{Equal contribution.}\def\thefootnote{\arabic{footnote}}

\section{Introduction}
Processing-in-memory (PIM)~\cite{modern_primer, pim0, pim1, enabling_pim} is a computing paradigm that places compute units in the memory chip to avoid moving data between the memory and the CPU cores.
This paradigm has been shown to offer significantly higher memory bandwidth and lower memory access latency for a {wide variety} of applications, including graph processing~\cite{pim_graph, graphide, nai2017, ahn.pei.isca15, pim_application8}, machine learning~\cite{gomez2022machine, gomez2023evaluating, gomez2022experimental, ke2021near, deep_hash, boroumand2018, boroumand2021google, pim_application0, pim_application1, pim_application2, pim_application3, pim_application4, pim_application5, pim_application6, newton}, database operations~\cite{axdimm_db, prim, ambit, boroumand2021icde, lim2023design, pim_application7}, sparse linear algebra~\cite{sparsep, pim_application8}, stochastic computing~\cite{SC}, {climate modeling~\cite{singh2020nero, singh2021accelerating}, stencil computations~\cite{denzler2023casper}, mobile applications~\cite{boroumand2018},} and bioinformatics~\cite{BLASTS, PIMQuantifier, PIMAssembler, PIMAligner, diab2022hicomb, kim2018, singh2021fpga}. 
The UPMEM PIM system~\cite{upmem, upmem1} is the first commercially available PIM hardware with general-purpose cores embedded in the DRAM chip.
Prior works show that the UPMEM PIM system can benefit numerous applications~\cite{prim, skyline, sparsep, DNA_upmem, gomez2022machine, gomez2023evaluating, gomez2022experimental, diab2022hicomb, diab2023framework, case_study, nider2020, lavenier2020, abecassis2023gapim, item2023transpimlib, lim2023design, gomez2022benchmarking, gupta2023homomorphic}.
%Many  
Several other similar PIM architectures have also been presented and prototyped~\cite{axdimm_db, FIMDRAM, lee2021hardware, ke2021near, lee2022isscc, niu2022isscc}. 

Programming a real PIM system is a challenging task due to the complexities involved.
For example, the UPMEM PIM system requires programmers to distribute data across the DRAM banks, launch PIM kernels on the PIM cores, manage the transfer of data between the DRAM banks and the PIM cores, and orchestrate the execution of multiple PIM threads on each PIM core~\cite{upmem-sdk}.
This task requires deep knowledge of PIM hardware and system architecture as well as proficiency in low-level APIs, which presents a steep learning curve. %\jgl{This citation is misplaced. It's not for the steep learning curve, but for UPMEM programming. Put it earlier}. \jfc{done}
Suitable library, programming model, compiler, and tool support is crucial for {adopting} PIM in real-world systems, as previously {discussed} in the literature~\cite{enabling_pim, modern_primer}.

To facilitate the adoption of PIM in real-world systems, we propose a high-level programming framework called SimplePIM.
SimplePIM abstracts the complexities of PIM hardware, supports multiple important applications such as histogram and K-means, and delivers high performance.%\jgll{Maybe give examples?} %\jgl{What do you mean by "essential applications"?} \jfc{I meant important applications, now changed.} 

The PIM systems targeted by our framework resemble distributed systems, where each PIM core has exclusive access to a memory region. However, {a} distributed system typically {involves} individual machines connected via a network and {needs} to handle node crash failures~\cite{Spanner, mapreduce, Paxos, Raft, ZooKeeper, Bully} or even {Byzantine} failures~\cite{byzantine0, byzantine1, byzantine2, lamport1982byzantine}. Many distributed systems use voting protocols to elect coordinator nodes for system management~\cite{Bully, Paxos, Raft, ZooKeeper}. However, failure handling should not be a concern for PIM since the PIM cores are not constantly failing and re-joining the system. In PIM systems, the host CPU is responsible for coordinating the PIM cores and handling communication between them. SimplePIM leverages the power of the central host to manage the entire system, including bookkeeping {of} the framework {metadata} and merging {of} intermediate results from PIM cores. Additionally, the host CPU facilitates communication with the outside world, such as network or I/O operations. %\jgl{This paragraph needs citations.} \jfc{done}

To support PIM systems, SimplePIM provides iterators such as \texttt{map}, \texttt{reduce}, and \texttt{zip}. These iterators are commonly found in programming languages (e.g., Python, Haskell) and distributed frameworks (e.g., MapReduce~\cite{mapreduce}, Spark~\cite{Spark}). Their purpose %of these iterators 
is to separate the application logic from the parallel decomposition of work across cores and threads. %  \jgl{Cite them} \jfc{made connection to built-in iterators and distributed systems, haven't found papers introducing iterator.}

To facilitate communication of data between the host CPU and the PIM cores, SimplePIM also provides \texttt{broadcast}, \texttt{scatter}, and \texttt{gather} collective communication techniques that involve the host CPU as the root node.
Communication primitives among the PIM cores, such as \texttt{allreduce} and \texttt{allgather}, are also available. %\jgl{Please "emph" or "texttt" terms: map, reduce, zip, broadcast, scatter, gather, allreduce, allgather...} \jfc{ok}
The communication interface is similar to the message-passing paradigm used in MPI~\cite{mpi}.
However, unlike MPI, which is homogenous and fully distributed, in SimplePIM, the host CPU always plays the unique role of the root node in managing the entire system and merging the intermediate results from different PIM cores.

We implement and evaluate SimplePIM on a real PIM system, UPMEM~\cite{upmem}, with six different applications: reduction, vector addition, histogram, linear regression, logistic regression, and K-means.
These applications have previously been implemented on UPMEM~\cite{prim, gomez2022machine, gomez2023evaluating, gomez2022experimental}, providing a baseline for comparing performance, correctness, and code complexity.
SimplePIM offers a {programmer}-friendly interface and requires $4.4\times$ fewer lines of code, on average, compared to {the best} existing open-source {hand-optimized} implementations.
In addition, we apply several code optimizations to tailor our SimplePIM implementation to the underlying hardware, making it suitable for UPMEM.
Our evaluation results show that SimplePIM performs similarly to hand-optimized implementations in three applications, and outperforms them in the remaining three, despite its general-purpose design. Specifically, for vector addition, logistic regression, and K-means, SimplePIM performs 1.10$\times$, 1.17$\times$, and 1.37$\times$ faster than {the best prior hand-optimized implementations} in weak scaling tests and 1.15$\times$, 1.22$\times$, and 1.43$\times$ faster in strong scaling tests. %\jgl{Give actual numbers. \jfc{done}

Our main contributions are:

\begin{itemize}

 \item We design {and introduce} SimplePIM, the first {high-level programming} framework tailored to improve programming productivity in {general-purpose} PIM architectures. 
 {We open-source SimplePIM at \url{https://github.com/CMU-SAFARI/SimplePIM} to enhance programming accessibility and {aid} the adoption of PIM systems.}

 \item We {implement} SimplePIM on the UPMEM PIM architecture and {develop} six PIM workloads (reduction, vector addition, histogram, linear regression, logistic regression, and K-means) using it. SimplePIM provides a significant reduction in {the number of lines of} code ranging from 66.5\% to 83.1\%, %resulting in 3$\times$ to 5.92$\times$ productivity improvement 
 {i.e., productivity improvements of 2.98$\times$ to 5.93$\times$} 
 compared to {the best prior hand-optimized open-source implementations} written using the UPMEM SDK.

 \item We {explore and implement} performance optimization techniques both in general and specific to SimplePIM. The SimplePIM implementation of the {evaluated} workloads performs similarly to {hand-optimized} implementations in three applications and achieves a speedup of 1.10$\times$-1.43$\times$ in the other three applications.

\end{itemize}

%\jgl{We need a list of contributions here.} \jfc{done}

%We intend to open-source SimplePIM and share all the application experiments in this paper to enhance programming accessibility and promote PIM system adoption.

%% file: secs/2-Background.tex
\section{Background}
\label{sec:bg}

%\jgl{We have related work section at the end of the paper. Let's just call this section "Background".} \jfc{done}

%\jgl{There is an arrow missing between host CPU and PIM DIMMs.} \jgl{This huge caption is not good. Captions should be short. All explanations here should go into the text.} \jfc{done}

Numerous {real} PIM architectures have been introduced that aim to bring compute units closer to memory~\cite{upmem, upmem1, axdimm_db, FIMDRAM, lee2021hardware, ke2021near, lee2022isscc, niu2022isscc}.
{For example,} AxDIMM\cite{ke2021near} {places an FPGA near DRAM ranks} to accelerate recommendation systems~\cite{ke2021near} and database operators~\cite{axdimm_db}. 
{FIMDRAM~\cite{FIMDRAM} features vector processing units near the banks of High Bandwidth Memory (HBM). FIMDRAM is} specifically designed for deep learning applications. 
{SK Hynix AiM~\cite{lee2022isscc} is also designed for deep learning applications. AiM places vector processing units near the banks of GDDR6 memory. 
Alibaba HB-PNM~\cite{niu2022isscc} glues together one layer of DRAM and one logic layer with processing elements designed to accelerate recommendation systems.}

Our programming framework implementation in this work uses the commercially available UPMEM PIM system~\cite{upmem}. 
{Two major characteristics distinguish UPMEM from other real PIM systems: 
(1) it integrates general-purpose processing cores in DRAM chips, and 
(2) it is the only commercially available PIM architecture (as of September 2023).}

\begin{figure}[h]
    \centering
    \includegraphics[width=\linewidth]{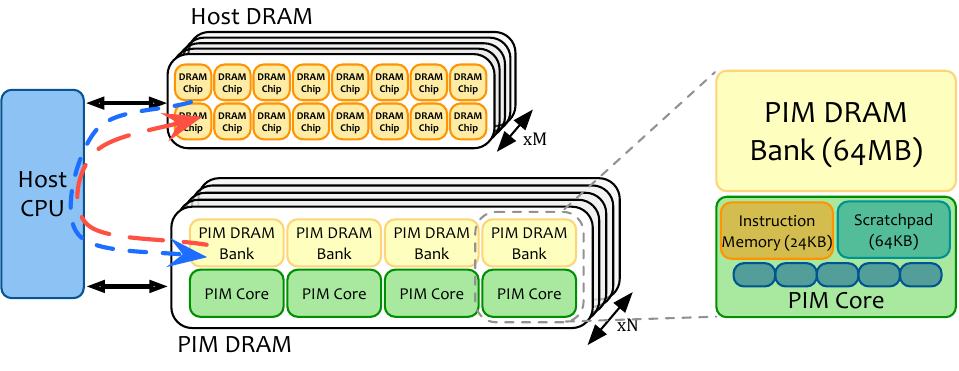}
    %\vspace{-20pt}
    \caption{The UPMEM PIM Architecture}\label{fig:upmem} 
    %\vspace{-12pt}
\end{figure}

%Fig.~\ref{fig:upmem}\jgl{I added a space between Fig. and 1. Make sure it's the same everywhere.} \jfc{ok}
Previous studies extensively {investigate} the UPMEM system~\cite{prim, skyline, sparsep, DNA_upmem, gomez2022machine, gomez2023evaluating, gomez2022experimental, diab2022hicomb, diab2023framework, case_study, nider2020, lavenier2020, abecassis2023gapim, item2023transpimlib, lim2023design, gomez2022benchmarking, gupta2023homomorphic}. 
Fig.~\ref{fig:upmem} provides a simplified view of the {first-generation} UPMEM architecture from a programmer's perspective. 
The UPMEM hardware is connected to the host system {via} DRAM DIMM connections. 
A state-of-the-art UPMEM server contains {up to} 20 PIM-enabled DIMMs, each with 2 ranks of 8 PIM chips. 
Each PIM chip has eight 64MB DRAM banks, with a programmable PIM core, a 64KB scratchpad memory, and 24KB instruction memory coupled to each bank. 
In total, there are {up to} 2,560 PIM cores. 
Data transfers between the scratchpad memory and the DRAM bank occur through explicit data transfer commands, with a maximum bandwidth of 800MB/s per bank. %Our system comprises 2,432 PIM DRAM banks, providing 
{The entire PIM system provides} a maximum bandwidth of {2TB/s for all PIM cores.} 

The PIM cores operate at 450 MHz and feature an 11-stage pipeline. They can perform one integer addition/subtraction per cycle and 32-bit integer multiplication/division in, {at most,} 32 cycles when the pipeline is fully utilized. However, floating-point operations may take tens to 2000 cycles to complete, as explained in~\cite{prim}. The 2,560 independent PIM cores operate in parallel, providing a peak compute {throughput} exceeding 1 {TOPS (Tera operations per second)}. 
There is no direct communication mechanism among different PIM cores in hardware. 
All communication occurs through memory {transfers} between the host DRAM and the PIM DRAM.

Writing a functionally correct PIM program for a system like the UPMEM PIM architecture is challenging, and optimizing code for performance requires programmers to have a deep understanding of the PIM hardware. 
To program PIM cores, {programmers} must manually manage {the} scratchpad memory to ensure good performance. 
Data {transfers} between the scratchpad memory and {the corresponding} PIM DRAM bank must be 8-byte aligned and not exceed a 2,048-byte limit. 
For {multithreading, the UPMEM SDK~\cite{upmem-sdk}} provides barrier synchronization, handshakes, and mutexes. At least 11 software threads are required to fully utilize the pipeline~\cite{gomez2022benchmarking, prim}. 
On the {host CPU side, programmers} allocate/reallocate the number of PIM cores, load the PIM binary, and explicitly launch the PIM program using commands provided in the UPMEM SDK~\cite{upmem-sdk}. Communication between {the} PIM cores and the host CPU is enabled by gather/scatter, broadcast, parallel and serial transfer commands, {all of} which have different performance implications and data alignment constraints. 

\textbf{Our goal} in this work is to design a high-level programming framework {for PIM architectures} that abstracts these hardware-specific complexities and provides a clean yet powerful interface for ease of use and {high} program performance.

%\jgl{All this paragraph should go to the final related work section. The background section should just provide information that is necessary to understand the paper. These examples of UPMEM works are not necessary to understand the paper.} \jinfan{moved the paragraph to related work}

%% file: secs/3-SmallTable.tex
\section{The SimplePIM Programming Framework}

A software framework is a universal and reusable software environment that provides a standard abstraction to build and deploy applications.
Our software framework, SimplePIM, leverages the massive memory bandwidth and parallelism offered by PIM hardware to operate on arrays of arbitrary size and dimensions.
Similar abstractions are widely used in distributed system frameworks (e.g., Spark~\cite{Spark}, MapReduce~\cite{mapreduce}). 

SimplePIM offers three key interfaces to support PIM systems like UPMEM. 
The {\emph{management interface}} (Section~\ref{sec:management_interface}) %\jgll{Section X, capitalized. Revise everywhere.}
stores {metadata} for the PIM-resident arrays, which can be accessed by the {programmer} and other parts of SimplePIM as needed. 
The {\emph{communication interface}} (Section~\ref{sec:comm_interface}) provides abstractions for both Host-PIM and PIM-PIM communication patterns. 
These patterns are similar to communication patterns in other distributed frameworks such as MPI~\cite{mpi}, %\jgll{Cite}
{which makes it easier for SimplePIM to be adopted}. 
{The \emph{processing interface}} (Section~\ref{sec:processing_interface}) leverages PIM's high memory bandwidth and parallelism to execute \texttt{map}, \texttt{reduce}, and \texttt{zip} iterators on PIM arrays. 
{Programmers} can combine these iterators to implement many widely-used workloads ranging from simple vector addition to complex machine learning model training. Similar to SimplePIM, systems like Spark~\cite{Spark} provide {programmers} with some data transformation operations and iterators for processing data. 
%\jgl{For each of them, in parentheses, refer the section.} \jfc{done. for processing interface, I don't know if I should mention other iterators could be added upon needs} \jgll{Maybe... but you also need to say why you didn't include them now, e.g., no real workloads available on the target PIM architecture for comparison.}

%\jgl{See my comment in the caption of Figure 4. We need to start this section explaining our design decisions. Why we created SimplePIM the way we did? Why do we have map, reduction and zip? Why don't we have other primitives (e.g., filter)?} \jfc{I tried to explain the design decisions above, please check if that makes sense.}

\subsection{Management Interface} \label{sec:management_interface}

%\ie{You have plenty of space here. Maybe make the font smaller and add details such as how you represent types and distribution.}\jgl{Please make all captions much shorter. 2 lines are enough. Any other explanations should go into the text of the paper.} \jfc{removed the figure since not enough space and the concept is relatively simple} \jgll{Maybe you can put it back after commenting out all comments.}

The SimplePIM Management Interface provides three main {functions}: \texttt{lookup}, \texttt{register} and \texttt{free}. 
These APIs enable tracking of PIM-resident data, data allocation and data deallocation in the form of continuous arrays. This management is centralized and takes place on the host CPU.

The management interface defines and uses two data structures. 
{First,} the \texttt{array\_meta\_data\_t} struct contains several fields that describe the PIM-resident array. These {fields are} the ID of the array, its length, data type, and the physical address of its data in {the PIM DRAM}. 
{Second,} the \texttt{simple\_pim\_management\_t} struct is responsible for managing all the PIM-resident arrays registered by the programmer. 
It contains an array of \texttt{array\_meta\_data\_t} structs, along with other hardware information such as the number of PIM cores.

\paragraph{Lookup} The \texttt{lookup} function {retrieves} the struct of \texttt{array\_meta\_data\_t} containing all relevant information of an array from the management unit, based on its unique ID. 
This function is {used} by both the communication and processing interfaces to seamlessly access and manipulate the array with a single ID provided by the {programmer}. %\om{Double check all code.}

\begin{lstlisting}[style=CStyle]
array_meta_data_t* simple_pim_array_lookup(const char* id, simple_pim_management_t* management);
\end{lstlisting}

\paragraph{Register} The purpose of the \texttt{register} function is to register the {metadata} of an array in the management unit. 
Typically, the function is called by the processing and communication interfaces when a new output array is created. The {programmer} provides a unique ID when calling the interfaces, and SimplePIM determines other relevant {metadata and registers} the output array properly. 
\begin{lstlisting}[style=CStyle]
void simple_pim_array_register(array_meta_data_t* meta_data, simple_pim_management_t* management);
\end{lstlisting}

\paragraph{Free} The ID is removed from the management unit, indicating that the array with that ID is no longer available for processing or communication.

\begin{lstlisting}[style=CStyle]
void simple_pim_array_free(const char* id, simple_pim_management_t* management);
\end{lstlisting}

%\jgl{No idea what are lookup, register and free. You need to define things. You cannot assume any background that you are not giving in the background section. If these three are common terms in high-level programming frameworks, maybe you can have a subsection about high-level programming in the background section.} \jfc{explained via API}

%\jgl{Don't we have APIs for register, lookup and free? We can have the function declarations same as we have in the next two subsections.}\jfc{done}

%\jgl{All figures of the paper are very ugly. Colors and those thick black borders are not nice. We have much better figures in other papers, for example, in all our UPMEM papers. Please make the figures nicer. They are the first thing that reviewers are going to see.} \jfc{tried to improve, please give some feedback if the figures are ok.}

\subsection{Communication Interface}\label{sec:comm_interface}

The SimplePIM Communication Interface serves as a comprehensive solution for handling communication between the host CPU and PIM cores, and among PIM cores. 
This interface effectively manages the complexities of data transfer {alignment, address calculation}, and different PIM communication commands so that programmers need not worry about them.
%\jgl{Space missing. Fix everywhere.} \jfc{done, this reference no more valid} 

To support host-PIM communication, SimplePIM provides three functions. 
The \texttt{broadcast} {function sends the same array to all PIM cores}. 
{The \texttt{scatter} function divides a host array in equal-sized chunks and distributes them across the PIM DRAM banks. 
The \texttt{gather} function {reassembles} the scattered chunks into a host array.}

To support communication among PIM cores, SimplePIM includes \texttt{allreduce} and \texttt{allgather} functions used in a variety of applications (e.g., machine learning).

%\jgl{I think we should use a different font type for all these primitive names everywhere in the paper. Maybe use emph or maybe use texttt -- I think I prefer the latter.} \jfc{I will use texttt throughout the text}

\begin{figure}[h]
\centering
    %\vspace{-12pt}
    \includegraphics[width=\linewidth]{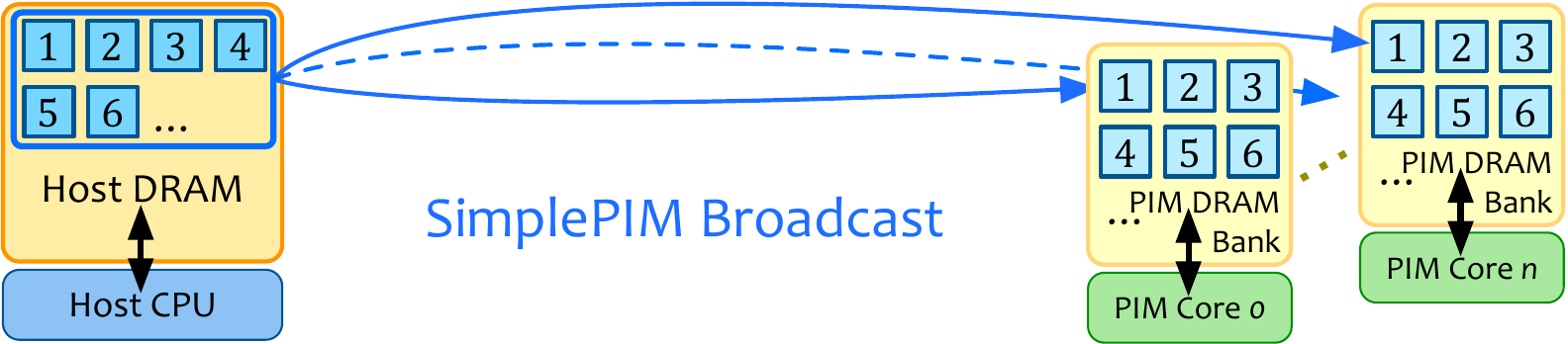}
    %\vspace{-20pt}
    \caption{SimplePIM Broadcast {function} (example with two PIM cores)} \label{fig:comm_bc}
    %\vspace{-12pt}
\end{figure}

\paragraph{Host-to-PIM {Communication}: SimplePIM Broadcast} The \texttt{broadcast} function in SimplePIM transfers a host array to all PIM cores in the system, ensuring that all PIM cores have a local copy of the data, as shown in Fig.~\ref{fig:comm_bc}. It then registers the ID of the array with the management interface so that it can be referred to by other functions in the SimplePIM interface. This function can be useful for initializing data or for distributing data that needs to be accessed by all PIM cores. 

In the \texttt{broadcast} function, the \texttt{arr} variable is the source array of the communication on the host side and the \texttt{type\_size} variable represents the size of a single element for the array. 
%The same naming convention is used for other primitives in the SimplePIM framework.

\begin{lstlisting}[style=CStyle]
void simple_pim_array_broadcast(char* const id, void* arr, uint64_t len, uint32_t type_size, simple_pim_management_t* management);
\end{lstlisting}
%\jgl{What are all the arguments of this function? We need to explain id, arr, len, type\_size...} \jfc{explained in texts}

%\jgl{We need to connect this explanation with the figures. We can use circled numbers to do so. We have plenty of space to improve the paper. We can have individual figures for each of these communication APIs.} \jgl{These figures need to be explained in the text. You should used circled numbers to clearly connect the figures and the text. The figures should be improved: there is a lot of white/empty space, but very small numbers inside those blue boxes.} \jfc{one figure for each primitive, explained in texts, figure improved}

\begin{figure}[h]
\centering
    %\vspace{-12pt}
    \includegraphics[width=\linewidth]{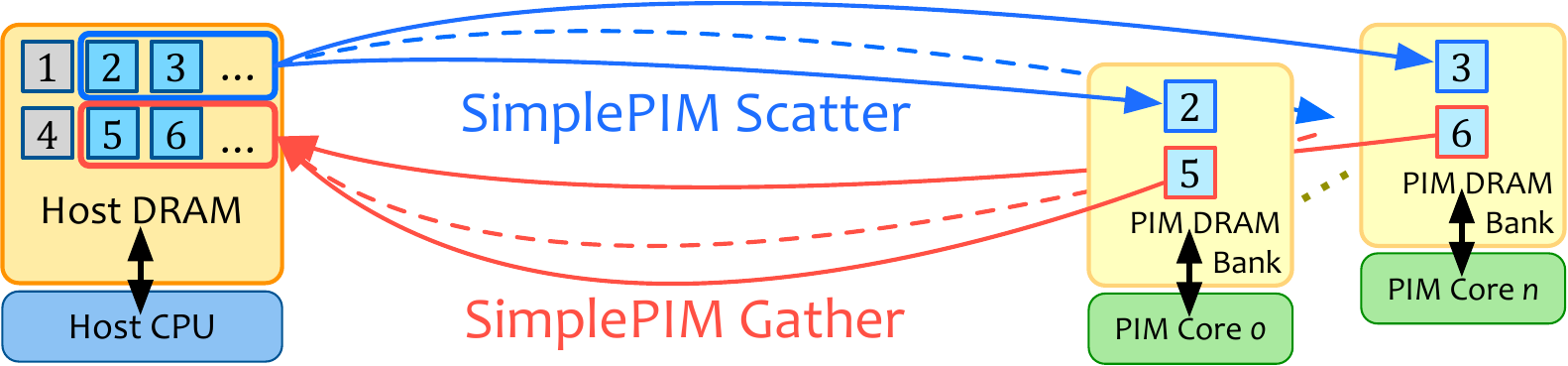}
    %\vspace{-20pt}
    \caption{SimplePIM Scatter {and} Gather {function} {(example with two PIM cores)}} \label{fig:comm_scatter_gather}
    %\vspace{-12pt}
\end{figure}

\paragraph{Host-to-PIM {Communication}: SimplePIM Scatter} The \texttt{scatter} function is designed to take an array located on the host {DRAM} and divide it into chunks that are distributed to each PIM core's {DRAM bank} (refer to Fig.~\ref{fig:comm_scatter_gather}). 
This division is performed almost evenly, while taking {into account} the PIM system's alignment constraints. 
For example, in the UPMEM system, data transfers must be aligned to eight bytes. %\jgl{This is implementation specific. I think we can just talk about this in Section 4.} \jfc{"divide almost evenly"  is the semantic of simple pim scatter, we could maybe still talk about it here.}. 
Once this division is complete, the \texttt{scatter} function registers the ID of the {destination} array {in the PIM DRAM banks (\texttt{id})} with the management interface. %\om{Isn't the array already registered?}.
{As in the \texttt{broadcast} function, \texttt{arr} and \texttt{type\_size} are, respectively, the source array and the array element size.}

\begin{lstlisting}[style=CStyle]
void simple_pim_array_scatter(char* const id, void* arr, uint64_t len, uint32_t type_size, simple_pim_management_t* management);
\end{lstlisting}

\paragraph{{PIM-to-Host Communication}: SimplePIM Gather} The \texttt{gather} function is the counterpart of the scatter function, serving to reassemble a scattered array, as shown in Fig.~\ref{fig:comm_scatter_gather}. It works by taking an identifier that corresponds to the scattered array and retrieving the relevant information through the memory management interface. Using this information, the \texttt{gather} function retrieves the split portions of the array from each PIM core's {DRAM bank}, collects them, and reassembles the original array on the host. Finally, the \texttt{gather} function returns a pointer to the gathered array. 

\begin{lstlisting}[style=CStyle]
void* simple_pim_array_gather(char* const id, simple_pim_management_t* management);
\end{lstlisting}

\begin{figure}[h]
\centering
    %\vspace{-12pt}
    \includegraphics[width=\linewidth]{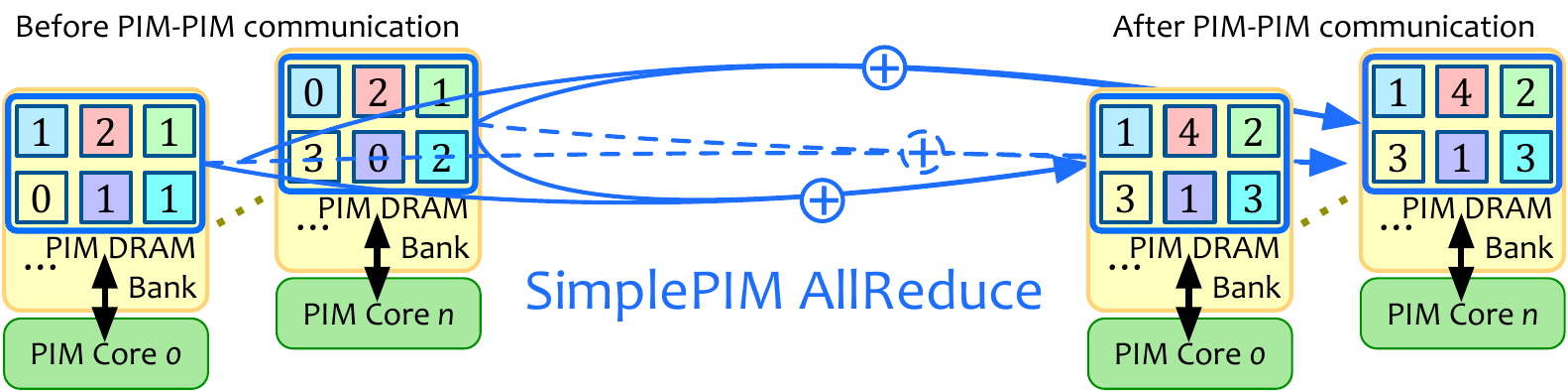}
    %\vspace{-20pt}
    \caption{SimplePIM AllReduce {function} {(example with two PIM cores)}} \label{fig:comm_allreduce}
    %\vspace{-12pt}
\end{figure}

\paragraph{PIM-PIM {Communication}: SimplePIM AllReduce} The \texttt{allreduce} function accepts %an array ID that is of 
{arrays} equal length across all PIM cores. 
To execute the \texttt{allreduce} operation, the {programmer} specifies an accumulative function {(i.e., the reduction operation)} and registers it as a function \texttt{handle}. 
SimplePIM then combines the arrays in place based on the {programmer}-defined function. 
Fig.~\ref{fig:comm_allreduce} shows an example {where the accumulative function is an addition}. 
{\texttt{allreduce} is often used for algorithm synchronization, for example} in machine learning applications. {Section~\ref{sec:processing_interface} details how a programmer creates a function \texttt{handle}.}

\begin{lstlisting}[style=CStyle]
void simple_pim_array_allreduce(char* const id, handle_t* handle, simple_pim_management_t* management);
\end{lstlisting}

\begin{figure}[h]
\centering
    %\vspace{-8pt}
    \includegraphics[width=\linewidth]{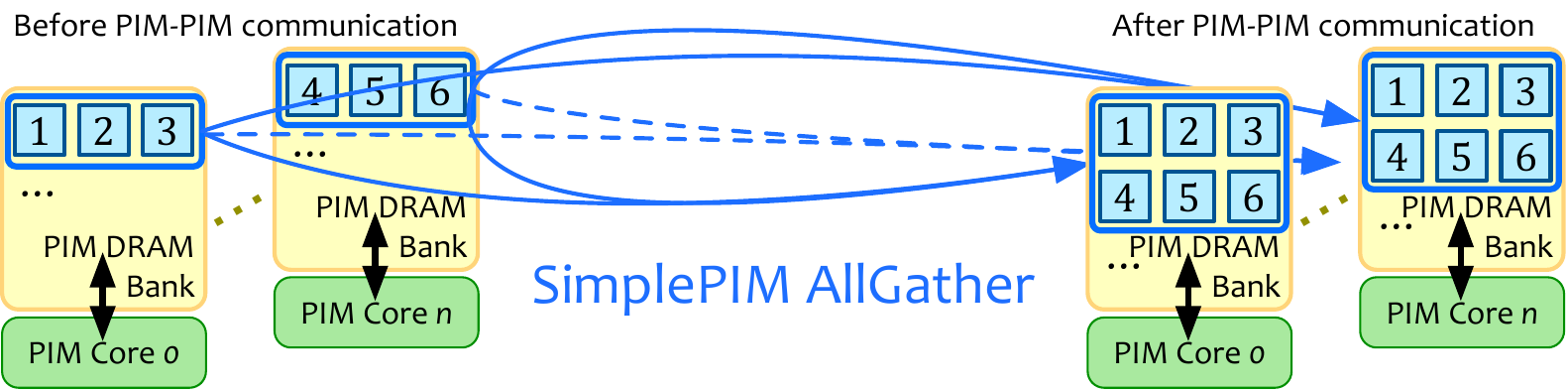}
    %\vspace{-20pt}
    \caption{SimplePIM AllGather {function} {(example with two PIM cores)}} \label{fig:comm_allgather}
    %\vspace{-14pt}
\end{figure}

\paragraph{PIM-PIM {Communication}: SimplePIM AllGather} The SimplePIM \texttt{allgather} function retrieves sections of an array from various PIM cores, combines them, and distributes the complete array to all PIM cores, as shown in Fig.~\ref{fig:comm_allgather}. This results in a new array with a unique identifier {\texttt{new id}}.

\begin{lstlisting}[style=CStyle]
void simple_pim_array_allgather(char* const id, char* new_id, simple_pim_management_t* management);
\end{lstlisting}

\subsection{Processing Interface} \label{sec:processing_interface}

% \ie{The listings in this section are missing the simple\_pim prefix.} \jfc{changed}

The SimplePIM Processing Interface provides iterators that enable the creation of a new array of data after performing operations on an existing array within the PIM system. These iterators can be called from the host and are parallelized automatically by the framework to execute across the PIM cores and threads. %, which leads to enhanced performance. 
Once executed, the resulting output of the iterator is registered via the management interface for future reference.

%After examining various PIM-suitable applications that have been implemented and benchmarked on real systems~\cite{upmem_benchmark, gomez2022machine, gomez2023evaluating, gomez2022experimental}, we have identified common patterns and have decided to support map, reduce, and zip during our first release. This decision is reflected in Fig.~\ref{fig:ops}. Furthermore, we plan to extend our support to other processing primitives, such as fold, in the future.

%\jgl{This last sentence makes no sense here. Why are we only implementing map, reduction and zip? It must be for a reason. We have to explain our reasons to design SimplePIM the way we did. With this sentence here, we are just shooting ourselves in the foot: we are giving bad reviewers very good ammunition to kill our paper.} \jfc{I removed the last sentence, the iterators are essential for our target applications, that's the most important (original figure removed)}

\paragraph{Creation of a Function Handle}

In {some PIM systems (such as UPMEM)}, PIM functions are loaded as independent binaries from the main program, compiled using a different compiler that targets the underlying PIM instruction set architecture. This separation ensures that CPU code does not reference PIM functions directly, although the CPU code must be able to pass PIM functions as inputs to SimplePIM's iterators. To facilitate this, the \texttt{create\_handle} function reads a file containing a PIM function, compiles the function, and provides a handle to the CPU that can be passed as an input to the iterators. The \texttt{transformation\_type} argument specifies which iterator the handle is for. 
The {programmer}-defined functions can execute with a context: the \texttt{data} of size \texttt{data\_size} is {broadcast} to all PIM cores, and the {programmer}-defined functions receive this data to aid their executions. For {example}, {for the linear regression workload}, the {programmer}-defined function requires model weight data to compute gradients, and this data is provided as context via the \texttt{data} argument.
%\ie{Update the handle to also take in the function's constant data and its size (instead of using the start function).}
% Some table transformations take functions as input. For example map :: (a \textrightarrow b) \textrightarrow [a] \textrightarrow [b] requires a map\_func :: (a \textrightarrow b). In SmallTable, the user puts necessary functions in a separate file and generates a function handle in the main program using 
\begin{lstlisting}[style=CStyle]
handle_t* simple_pim_create_handle(const char* func_filepath, uint32_t transformation_type, void* data, uint32_t data_size);
\end{lstlisting}

\paragraph{Array Map}

\begin{figure}[h]
\centering
    %\vspace{-8pt}
    \includegraphics[width=0.8\linewidth]{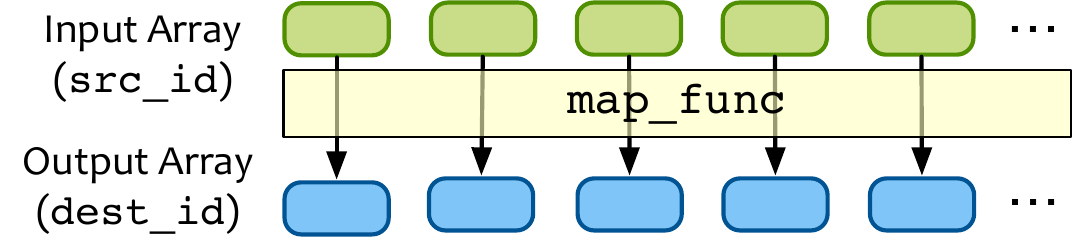}
    %\vspace{-10pt}
    \caption{SimplePIM Array Map {function}} \label{fig:map}
    %\vspace{-10pt}
\end{figure}

The \texttt{simple\_pim\_array\_map} iterator takes a registered input array and a function \texttt{handle}, applies the \texttt{map\_func} function to every element in the data array, and generates a new output array with \texttt{dest\_id}, as Fig.~\ref{fig:map} shows.

\begin{lstlisting}[style=CStyle]
void simple_pim_array_map(const char* src_id, const char* dest_id, uint32_t output_type, handle_t* handle, simple_pim_management_t* management);
\end{lstlisting}

%\ie{This and the next two figures can be made smaller to save space.} \jfc{input up, output down, make arrays clearer for zip.}

\paragraph{Array Reduction}
The \texttt{simple\_pim\_array\_red} iterator processes each element in an input array {\texttt{src\_id}}, calculates an index {to an output array \texttt{dest\_id} and performs a reduction onto the indexed output array element}. 
This operation is similar to the general reduction method proposed in the FREERIDE middleware~\cite{FREERIDE}. 
Later, MATE~\cite{MATE} %\om{Check references} 
implemented general reduction and demonstrated its superior performance compared to MapReduce in a multi-core system. SimplePIM's PIM reduction iterator is versatile enough to support various essential applications such as linear regression, K-means, and histogram {calculation}.

\begin{figure}[h]
\centering
    %\vspace{-5pt}
    \includegraphics[width=0.8\linewidth]{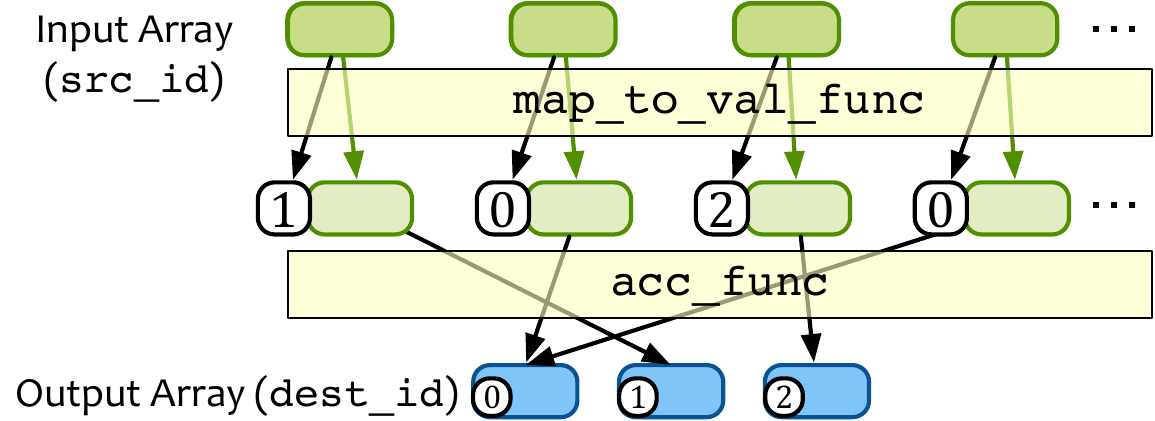}
    %\vspace{-10pt}
    \caption{SimplePIM Array Reduction {function}} \label{fig:reduce}
    %\vspace{-10pt}
\end{figure}

To perform the \texttt{PIM array reduction}, the {programmer} needs to define three functions. The \texttt{init\_func} initializes all entries in the output array. 
The \texttt{map\_to\_val\_func} function transforms an input element to an output element and determines the corresponding entry in the output array to accumulate the current output element. 
Finally, the commutative function \texttt{acc\_func} accumulates the output element {that results of the \texttt{map\_to\_val\_func} function onto} the corresponding entry in the output array. 
{Fig.~\ref{fig:reduce} shows the} usage of \texttt{map\_to\_val\_func} and \texttt{acc\_func} functions. 
In {the histogram calculation of an image, for example}, the \texttt{init\_func} initializes all histogram bins to 0. 
The \texttt{map\_to\_val\_func} computes for each {image pixel} which bin (i.e., an index in the histogram) it belongs to and returns 1. 
Finally, the \texttt{acc\_func} is a simple addition that increments the bin count by 1.

\begin{lstlisting}[style=CStyle]
void simple_pim_array_red(const char* src_id, const char* dest_id, uint32_t output_type, uint32_t output_len, handle_t* handle, simple_pim_management_t* management)
\end{lstlisting}

\begin{figure}[h]
\centering
    %\vspace{-8pt}
    \includegraphics[width=0.8\linewidth]{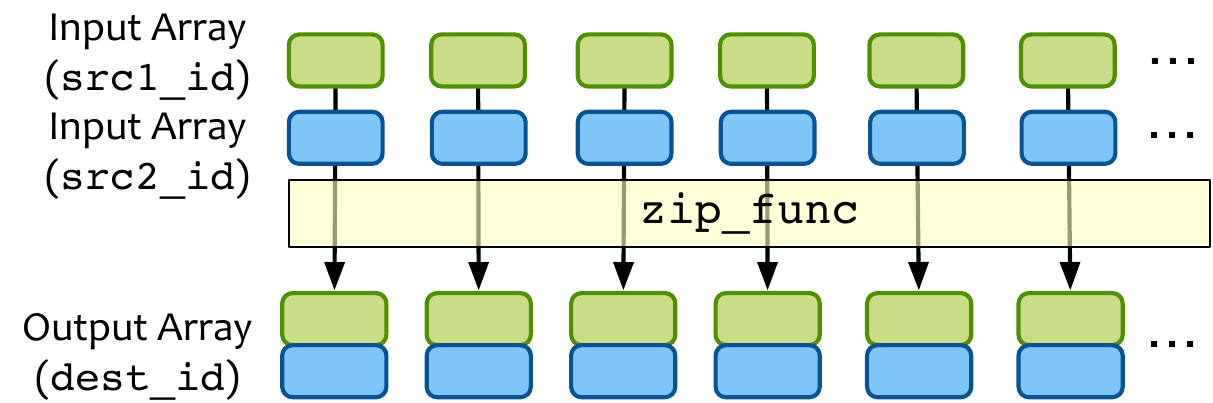}
    %\vspace{-10pt}
    \caption{SimplePIM Array Zip {function}} \label{fig:zip}
    %\vspace{-16pt}
\end{figure}

\paragraph{Array Zip}
The \texttt{simple\_pim\_array\_zip} iterator takes as input the IDs of two registered arrays that are of the same length. It then generates an {output} array that combines the elements of the two input arrays, {as Fig.~\ref{fig:zip} shows}. %and outputs the result. 
By allowing {programmers} to work with multiple arrays as inputs to the iterators, this function enables greater flexibility in data processing.

\begin{lstlisting}[style=CStyle]
void simple_pim_array_zip(const char* src1_id, const char* src2_id, const char* dest_id, simple_pim_management_t* management)
\end{lstlisting}

% In addition, the SimplePIM programming framework enables the programmer to register a start function for any processing iterator.
% This function aids in setting up the user-defined functions and is automatically called before the iterator processing logic.
% For example, in the case of linear regression, the user-defined function requires model weights.
% The start function is defined by the user to initialize the model weights before the regression logic is executed.
% \ie{I still don't like this. Model weights are typically copied from memory, not initialized by a function. Why do we need a function? Is there any use of the start function besides initializing constant data that is needed by the function? Can't we just pass data to the handle when creating it?}
% \ie{We agreed to omit the start function and pass data used by the function to the create handle call.}

% \ie{Why the model weights be provided as direct inputs to the iterators? Why is a start function needed for this? It gives the feeling that the iterators are not expressive enough. Also, whenever I hear "global variables", it makes me worried.} \jfc{changed the expression} \ie{Where is this info field provided in the listings?} \jfc{choose not to mention, confusing and less relevant}

%% file: secs/4-Implementation.tex
\section{UPMEM Implementation and Optimizations for SimplePIM}\label{sec:st-impl}

PIM {cores access} memory with higher bandwidth and lower latency {(than CPUs or GPUs)}, making this architecture more compute-bound, and highlighting the importance of code optimization for optimal performance.
In this section, we provide detailed information on how we develop and optimize the SimplePIM interface for the UPMEM system.
% While our extensive optimization efforts have yielded performance benefits for all applications using our framework, applications developed entirely from scratch may not have undergone thorough the same testing and optimization process, making them potentially less performant and error-prone. \ie{This can be misinterpreted as saying that our baselines are not strong, so better avoid it.}

\subsection{Communication Interface}
UPMEM provides basic serial commands for transferring data between a PIM core and the host, as well as parallel commands for transferring same-sized continuous data between the host and different PIM cores. The parallel transfer bandwidth increases with the number of PIM cores and can be orders of magnitude higher than {the serial transfer bandwidth}.

In order to use the parallel transfer commands, memory transfers must be aligned and equal-sized for all PIM cores. SimplePIM automatically pads the communicated arrays and determines the transfer size for the parallel commands based on the array size, data type, and alignment requirements, ensuring that no array element is split across PIM cores and all alignment requirements are met. This provides a clean interface to the {programmer}, as described in Section~\ref{sec:comm_interface}.

SimplePIM leverages parallel data transfers between {the UPMEM PIM cores} and the host to implement both PIM-PIM and {PIM-Host} communication primitives. Although previous works such as ABC-DIMM~\cite{ABC_DIMM} and DIMM-Link~\cite{DIMM_Link} {demonstrate} the potential benefits of hardware mechanisms for improving PIM-to-PIM communication, programmers {currently} do not have access to a physical {direct} communication channel among UPMEM PIM cores. In future PIM systems with more efficient communication among PIM cores enabled by hardware, SimplePIM could take advantage of these features to implement PIM-PIM communication more effectively.

\subsection{Processing Interface}

\subsubsection{Array Map} 
An input array on the host system can be evenly split among the PIM cores using communication primitives. When the \texttt{map} iterator is called from the host, all relevant PIM cores are simultaneously invoked to process their respective local arrays. \jfc{On each of these PIM cores, 12 PIM threads are launched by default, as at least 11 threads are required to fully utilize the pipelined in-order cores of the UPMEM architecture~\cite{prim, gomez2022benchmarking}. While we have chosen 12 threads as a convenient even number, the {programmers} have the flexibility to configure SimplePIM to run any desired number of threads.}
In SimplePIM, each PIM thread loads its assigned section of the input array from the PIM DRAM bank to the 64KB scratchpad in batches, maximizing the {scratchpad-to-DRAM} bandwidth. 
The thread then applies the \texttt{map\_func} to all elements of the input data and stores the results in the corresponding output address in batches as well.
% At the end of a map iterator, a barrier is used to ensure that the map operation is completed by all threads.
% This barrier synchronization is necessary since the user could invoke the communication interface (such as with \texttt{simple\_pim\_gather}) to retrieve the map results to host. The iterators must complete when the iterator calls are returned. The same barrier is also required for the \texttt{reduce} and \texttt{zip} iterators.
% % \ie{It's not clear why the barrier is necessary. The map iterator semantics do not require it. For example, if the map is followed by another map, the cores could proceed to executing the next map without waiting for each other.}
% % \jfc{explained in texts}
% \ie{This is not convincing. The barrier should be invoked at the beginning of the communication primitive in this case, not at the end of the iterator. Any reason you placed it at the end of map instead of at the beginning of gather?} \ie{Omit discussion of barrier}

\subsubsection{Array Reduction}\label{sec:reduction-impl} 
To perform array reduction, SimplePIM activates all PIM cores containing segments of the input and launches 12 PIM threads on each core. These threads operate on the input data in batches, and each PIM core generates its intermediate reduction result in parallel using the {programmer}-defined \texttt{acc\_func} PIM function. These intermediate results are then gathered to the host and combined using a host version of \texttt{acc\_func} with the help of OpenMP~\cite{OpenMP}.

% Notice that \texttt{acc\_func} needs to be executed on both the PIM cores and the host CPU.
% However, the host program does not know which handles it will create at compile time.
% For this reason, the host version of \texttt{acc\_func} is compiled at run time as a dynamically linked library during the handle creation process. % and loaded using the \texttt{dlopen} command provided by the operating system.
% The function must be named \texttt{acc\_func} so that the host can refer to it.
% Previously loaded versions of \texttt{acc\_func} are unloaded to avoid name conflicts.
% \ie{It's not clear how this works. The \texttt{func\_filepath} could be a variable not known at compile time. How would you extract a function name from it and call it?}
% \ie{Let's omit this to avoid confusion.}

% \ie{Maybe emphasize that the accum\_func was also used on the PIM side, which means there needs to be two versions of the function compiled for the host and PIM.}
% \ie{I think more clarification is needed here. How does the host CPU invoke the function if it is compiled at runtime? How does it know its name? Or are you assigning your own names? If yes, how do you manage multiple functions? Also, usually dynamically linked libraies are linked at load time, but here you need to link at runtime. Is there anything special you need to do to link at runtime?} \jfc{explained in texts}

Since an entry in the output array is accessed once per input element, a straightforward optimization is to keep the output array in the 64KB scratchpad {memory}, if it fits. 
SimplePIM offers two variants of {in-scratchpad} \texttt{PIM array reduction}: shared and thread-private accumulators. %\ie{I suggest referring to them a thread-private and shared accumulators. I think locks are distracting here.}

%\jgl{No huge captions please. You can use larger font in all plots. Also there are nicer color palettes than these "mustard yellow" and "fluorescent lime".} \jgl{Btw, why is this plot here and not in Section 5? Results should go into the results section.} \jfc{made caption smaller, the figure explains part of the optimization techniques so I thought it should go here..}

\paragraph{Shared Accumulator Reduction} {In this variant,} only one reduction output array is present in the scratchpad {memory}, and one lock is preserved per array entry. One thread initializes this array at the beginning with \texttt{init\_func}. Then, every thread works on its part of the input array and updates the global output array entries after having acquired the lock associated with {each} entry.

%% maybe I should not call it lock-free...
\paragraph{Thread-Private Accumulator Reduction} {In this variant,} every thread initializes its own local output array with \texttt{init\_func} and reduces its input segment to that local output array with the \texttt{map\_to\_val} and \texttt{acc\_func} functions. Once all threads {finish} the reduction, local outputs are merged in parallel in a ring-reduce manner with \texttt{acc\_func} and barriers.

These two implementations have a {tradeoff} between synchronization overhead and {scratchpad} capacity utilization. 
The framework automatically chooses an appropriate {in-scratchpad} reduction variant based on the array sizes and data types. 
We evaluate the {tradeoff} between these two variants in Section~\ref{sec:reduction-eval}.

\subsubsection{Array Zip}
The \texttt{zip} iterator is implemented with a lazy approach to minimize data copying.
When the zip iterator is called, the management interface stores the starting addresses, data type sizes, and a common length for the two arrays to be zipped, but does not physically combine them.
When a subsequent iterator is passed a lazily zipped array, {the management interface} extracts the addresses of the two contained arrays and copies the arrays in batches to the scratchpad. 
% \ie{I wouldn't say different versions of the iterators because that sounds like bad modularity. Instead, you can say that the iterators behave differently when they are operating on a zipped array.} \jfc{changed in texts}
Then, the batches are zipped in the scratchpad, and the iterator operation is performed.
This approach ensures that data is copied only once in the same loop, reducing both copying and looping overheads.
The management interface keeps track of lazily zipped arrays and determines the behavior of the iterators.
Currently, the laziness in our implementation is one level deep, which is sufficient for targeting the common case (i.e., multi-input map and reduce functions).
If the zip iterator is called with an already lazily zipped array as an argument, the arrays are streamed in batches, combined physically, and stored back to memory in batches.
Our experiments {(not {plotted} in the paper)} show that lazy zipping improves the performance of vector addition, for example, by more than 2$\times$.

\subsection{General Code Optimizations} We {implement} several code optimization techniques in SimplePIM that are essential for achieving high performance on UPMEM. 
{Some} open-sourced UPMEM {applications use only some} of these optimizations, making them potentially slower than SimplePIM {implementations}. 
{We list} some of the optimizations {SimplePIM employs}:

\begin{enumerate}
    \item \emph{Strength Reduction}: The UPMEM chip lacks native support for 32/64-bit integer {multiplication}, which must be emulated by runtime software and can take tens of cycles to complete. To overcome this challenge, SimplePIM strives to minimize the use of multiplications in the main loop. This is especially {important} for a general-purpose framework like SimplePIM, where the compiler may not always perform strength reduction automatically. Although the array data type is not known at compile time, the sizes of array elements are often powers of two. SimplePIM takes advantage of this fact by %handling power-of-two cases and 
    replacing multiplications with {shift operations in array offset calculations when the array size is a power of two}.
    
    \item \emph{Loop Unrolling}: We have found that loop unrolling can improve the performance of vector addition by up to 20\% on UPMEM. We attribute this performance gain to fewer loop counter increments and loop branches. However, {loop unrolling} can also increase the binary size, which may eventually not fit into the instruction {memory} of PIM {cores}. Our SimplePIM implementation for UPMEM uses limited unrolling depth.
    
    \item \emph{Avoiding Boundary Checks}: Many open-sourced UPMEM applications check array boundaries inside the main loop for convenience. In SimplePIM, we evenly pre-partition the work among threads and then process the trailing part of the array separately to avoid boundary checks. For example, we have experienced more than 10\% performance {degradation} due to boundary checks for the vector addition application.

    \item \emph{Function {Inlining}}:
    % In the initial implementation of SimplePIM, we observed serious performance problems due to the function call overhead in the main loops of the map and reduce iterators.
    % To tackle this problem, SimplePIM incorporates user-defined functions in a header file and directly includes them in the iterator code. This allows for seamless compilation during the handle creation process.
    % \ie{This is not clear. Do you mean to say that the iterators themselves that use each function are generated during the handle creation process? Please clarify.} \jfc{explained in texts} \ie{So the answer is yes?} \jfc{yes, I think}
    % By doing so, the compiler views the function body and function call as a single unit and can inline the function where it deems appropriate.
    SimplePIM inlines {programmer}-defined functions in the iterator code to avoid the function invocation overhead in the iterator loops.
    That is, at handle creation time, the functions are not compiled independently, but rather, the iterators that use the functions are also compiled.
    Compared to compiling iterators and functions separately, we have found that inlining improves the performance of vector addition by more than 2$\times$.

    \item Previous studies have shown that the performance of data transfers between the {PIM core's scratchpad memory and the corresponding DRAM bank} in UPMEM is highly dependent on the data transfer size~\cite{prim}.
    We {observe} that, in previously open-sourced implementations, programmers {fix} the {scratchpad-to-DRAM transfer} sizes for convenience, which results in suboptimal performance. 
    {An example is a linear regression implementation where the input dimensions are different for different datasets, which would require manual adjustment of transfer sizes.} 
    In contrast, SimplePIM automatically and dynamically adapts the {scratchpad-to-DRAM} transfer size to the input data size and type, achieving better performance while {freeing programmers from applying low-level optimizations and allowing them to focus on the application logic}. 
    % While many programmers use fixed-size buffers for data transfer, this approach can be problematic for complex applications such as linear regression.
    % We observed that, in previously open-sourced linear regression implementations, the buffer size needs to be manually adjusted with the input dimension to ensure correctness.
    % This is because the human developers pre-allocated a 2KB buffer on the scratchpad memory for data transfers to fully utilize the transfer bandwidth, but the size of a data point (which changes with feature dimension) needs to be divisible by the buffer size in their implementation to ensure correctness.
    % In contrast, a software framework like SimplePIM allows programmers to focus solely on application logic. Based on the input and output data sizes, the data transfer size is automatically determined for both correctness and performance.
    % \ie{Still not clear}
    % \ie{How is it determined? What's challenging about determining it?} \jfc{explained in texts}

\end{enumerate}

%% file: secs/5-Experiments.tex
\section{Evaluation}

\subsection{Benchmarks}

We evaluate SimplePIM using six {commonly-used} PIM-friendly applications: reduction, vector addition, histogram, linear regression, logistic regression, and K-means clustering. 
As a baseline for comparison, we use hand-tuned implementations from prior works~\cite{prim, gomez2022benchmarking, gomez2022machine, gomez2023evaluating, gomez2022experimental}. 
{The baselines for} reduction, vector addition, and histogram are from the PrIM benchmark suite~\cite{prim, gomez2022benchmarking, gomezluna2021repo}, while the {baselines for the three} machine learning algorithms, {i.e.,} linear regression, logistic regression, and K-means, {are} from~\cite{gomez2022machine, gomez2023evaluating, gomez2022experimental, gomezluna2023repo}. 
{These prior works develop} clean and high-performance {open-source codes} for benchmarking against CPU and GPU implementations. 
{As such, these codes serve as solid baselines for comparison with our} SimplePIM implementations.

\paragraph{Vector Addition} We implement \texttt{vector addition} in SimplePIM by zipping the input arrays and performing element-wise addition using the \texttt{map} iterator. SimplePIM automatically optimizes this operation by performing lazy zipping on the UPMEM device, {which results} in a high-performance implementation. 
We evaluate the {runtime} of \texttt{vector addition} for one million 32-bit integer elements per PIM core for weak scaling and 608,000,000 32-bit integer elements for strong scaling, similar to the reference {work that provides the baseline implementation}~\cite{prim, gomez2022benchmarking}.

\paragraph{Reduction} The \texttt{reduction} operation computes the sum of all elements in an input array. 
{In SimplePIM, we implement \texttt{reduction} using \texttt{PIM array reduction} with an output array of a single element (an accumulator).} 
We select the same number of {input} elements {for weak scaling and strong scaling} as for the \texttt{vector addition} workload.

\paragraph{Histogram} The \texttt{histogram} operation is implemented using \texttt{PIM array reduction}. We define a {programmer}-specific \texttt{map\_to\_val} function to compute the corresponding bin for each input element, and a simple addition is used to combine the element counts for each bin. We conducted experiments with 1,572,864 elements per PIM core for weak scaling and 956,301,312 for strong scaling, with the number of histogram bins set to 256 to ensure consistency with the reference {baseline implementation}~\cite{prim, gomez2022benchmarking}.

\paragraph{K-Means} The UPMEM PIM device currently emulates floating point operations in software, resulting in significantly slower performance {than integer operations}. To mitigate this issue, {our \texttt{k-means} benchmark employs} input data quantization to integers, following the approach outlined in \cite{gomez2022machine, gomez2023evaluating, gomez2022experimental}. {We conduct experiments} with 10 centroids and 10 feature dimensions, using 10,000 elements per PIM core for weak scaling experiments and 6,080,000 elements for strong scaling experiments. 

\paragraph{Linear Regression} To address the issue of slow floating point operations, we rely on the baseline approach proposed in \cite{gomez2022machine, gomez2023evaluating, gomez2022experimental}, which includes several versions of \texttt{linear regression} using various quantization techniques. For our experiments, we use the implementation that employs 32-bit integer operations with bit shifts to prevent integer overflow and underflow. To ensure a fair comparison, we verify that our SimplePIM implementation produces identical results to the baseline approach. We use a feature dimension of ten and generate 10,000 data points per PIM core for weak scaling tests, while for strong scaling tests, we generate a total of 6,080,000 data points, similar to those used in the {work that provides the baseline implementation~\cite{gomez2022machine, gomez2023evaluating, gomez2022experimental}}. 

\paragraph{Logistic Regression} To enable a fair comparison with the baseline approach \cite{gomez2022machine, gomez2023evaluating, gomez2022experimental}, we apply the same quantization technique used in \texttt{linear regression} to \texttt{logistic regression}. To minimize computational overhead, we adopt the Taylor series approximation of the sigmoid activation function \cite{sigmoid} {that the baseline~\cite{gomez2022machine, gomez2023evaluating, gomez2022experimental} uses}. However, since the runtime of the approximation depends on the input, we ensure a fair comparison by using the same inputs and initial model weights for both {the baseline code~\cite{gomez2022machine, gomez2023evaluating, gomez2022experimental} and the SimplePIM implementation}. We also verify that the exact same output is produced. The weak and strong scaling datasets have the same sizes as for \texttt{linear regression}.

\subsection{Productivity Improvement}

Efficiently implementing PIM kernels for these applications requires a significant amount of engineering effort. {We measure} the programming complexity by counting the lines of effective PIM-related code for each application. This excludes the common code for data loading from a file to the host main memory, host memory allocation, variable definition, and time measurements. We only take into account PIM-related data transfers and PIM kernel execution.
Table~\ref{table:lines_of_code} summarizes the lines of effective code saved by using SimplePIM for the six benchmarks.
The amount of coding is reduced by a factor of {$2.98\times$ to $5.93\times$} with SimplePIM.

\begin{table}[h]
  \centering
  \renewcommand{\arraystretch}{1.0}
\footnotesize
  \begin{tabular}{l|c|c|c}
    \toprule
    \hline
     & SimplePIM & {Hand-optimized} & LoC Reduction \\
    \hline\hline
    Reduction & 14 & 83 & {5.93}$\times$ \\
    \hline
    Vector Addition & 14  & 82  &  5.86$\times$ \\
    \hline
    Histogram & 21  & 114  & {5.43}$\times$  \\
    \hline
    Linear Regression & 48  &  157 & {3.27}$\times$ \\
    \hline
    Logistic Regression & 59 & 176 & {2.98}$\times$ \\
    \hline
     K-Means & 68  & 206  & {3.03}$\times$  \\
    \hline
    \bottomrule
  \end{tabular}
  \caption{{Lines of effective PIM-related code for each benchmark.} "LoC Reduction" stands for {SimplePIM's} reduction in lines of code {over hand-optimized baselines}.} % and "Reg." for regression.}
  \label{table:lines_of_code}
  %\vspace{-14pt}
\end{table}

%\jgl{Have a column for the abbreviations.}\jgl{No need to abbreviate "Application".}\jgl{Improvement: isn't "LoC reduction" better?} \jfc{added explanation for abbreviations in caption, is it ok?}

% \ie{In the rest of this section, I suggest swapping Listing 1 and Listing 2. Show the code before first, then the code after.} \jfc{done}

Productivity improvement is not only achieved by reducing the {lines of code}, but also by allowing {programmers} to write plain and easily understandable C code for an uncommon (PIM) architecture. 
Listing~\ref{hist_code_scratch} shows the code required to implement a {hand-optimized \texttt{histogram} operation} on the UPMEM {PIM architecture}. %\jgl{Bad writing. Say "using the UPMEM PIM programming framework" and cite it. Do not say "scratch", it reminds me a programming framework for kids.}. \jfc{rewritten} 
In this code, variable declarations, initialization, as well as code for initializing and computing the histogram are omitted. To program the PIM system, the programmer {is responsible for addressing {the data structures} (based on thread ID or \texttt{tasklet\_id}) and} 
must be familiar with architecture-specific instructions, such as {\texttt{mem\_reset} (line 5), \texttt{mram\_read} (line 14), and \texttt{mram\_write} (lines 26 \& 29)}.
%\jgl{Fix all these.} \jinfan{done}. 
The programmer needs to read the documentation carefully and understand the instructions {and their low-level properties} in detail. For example, the {\texttt{mram\_read} and \texttt{mram\_write} instructions, used for DRAM-scratchpad transfers,} have an 8-byte alignment requirement and a 2,048-byte transfer limit, and the programmer needs to handle larger transfers manually, as shown {in lines 28-30 of} Listing~\ref{hist_code_scratch}. Additionally, the code allocates a 2,048-byte buffer for {input data transfers (line 7)}. %since the input element size is fixed at 2, 4, or 8 bytes. 
For more complex applications with variable input element size, such as \texttt{linear regression}, the programmer needs to handle {data transfers} with greater care and effort.

%\jgl{What do you mean by "Scratch"?} \jfc{codes all written by human} \jgll{Say "Hand-written Code using UPMEM SDK"}

\begin{figure}[h]
\begin{lstlisting}[style=CStyle, breaklines=true, caption=Hand-optimized \texttt{histogram} code using {the} UPMEM SDK, label=hist_code_scratch, basicstyle=\scriptsize\ttfamily]
... // Initialize global variables and functions for histogram 
int main_kernel() {
  ... // Initialize variables and the histogram
  if (tasklet_id == 0)
    mem_reset(); // Reset the heap
  // Allocate buffer in scratchpad memory
  T *input_buff_A = (T*)mem_alloc(2048);
  
  for (unsigned int byte_index = base_tasklet; byte_index < input_size; byte_index += stride) {
    // Boundary checking
    uint32_t l_size_bytes = (byte_index + 2048 >= input_size) ? (input_size - byte_index) : 2048;     

    // Load scratchpad with a DRAM block
    mram_read((const __mram_ptr void*)(mram_base_addr_A + byte_index), input_buff_A, l_size_bytes); 

    // Histogram calculation
    histogram(hist, bins, input_buff_A, l_size_bytes/sizeof(uint32_t));
  }
  ...
  barrier_wait(&my_barrier); // Barrier to synchronize PIM threads
  ... // Merging histograms from different tasklets into one histo_dpu

  // Write result from scratchpad to DRAM
  if (tasklet_id == 0) {
    if (bins * sizeof(uint32_t) <= 2048)
      mram_write(histo_dpu, (__mram_ptr void*)mram_base_addr_histo, bins * sizeof(uint32_t));
    else 
      for (unsigned int offset = 0; offset < ((bins * sizeof(uint32_t)) >> 11); offset++) {
        mram_write(histo_dpu + (offset << 9), (__mram_ptr void*)(mram_base_addr_histo + (offset << 11)), 2048);
      }
  }
  return 0;
}

\end{lstlisting}
\end{figure}

Listing \ref{hist_code_sp} illustrates how SimplePIM makes it possible for {programmers} to implement \texttt{histogram} without resorting to any hardware-specific instruction or function calls. The code is straightforward and can be easily understood and implemented by any C programmer. The process involves defining application logic functions {(lines 2-14)} for the iterator in a separate file {(called \texttt{histo\_filepath} in this example)}, creating a handle {(line 17)}, and running the iterator with {only} two additional lines of code on the host side {(lines {20 \& 23})}. 
Similar to Listing~\ref{hist_code_scratch}, the code example in Listing \ref{hist_code_sp} does not show the host-side allocation and data transfers. 
% \ie{Should we also include the call to the iterator that uses these functions?} \jfc{we have not included host side API calls for human code also, we shouldn't include here for fairness (if then we should include both..)} \ie{I think we should include the iterator code without the transfer code because the iterator code is part of the application logic.}

\begin{figure}[h]
\begin{lstlisting}[style=CStyle, caption=SimplePIM \texttt{histogram} code, label=hist_code_sp, basicstyle=\scriptsize\ttfamily]
// Programmer-defined functions in the file "histo_filepath"
void init_func (uint32_t size, void* ptr) {
  char* casted_value_ptr = (char*) ptr;
  for (int i = 0; i < size; i++)
    casted_value_ptr[i] = 0;
}
void acc_func (void* dest, void* src) {
  *(uint32_t*)dest += *(uint32_t*)src; 
}
void map_to_val_func (void* input, void* output, uint32_t* key) {
  uint32_t d = *((uint32_t*)input);
  *(uint32_t*)output = 1;
  *key = d * bins >> 12;
}

// Host side handle creation and iterator call
handle_t* handle = simple_pim_create_handle("histo_filepath", REDUCE, NULL, 0);

// Transfer (scatter) data to PIM, register as "t1"
simple_pim_array_scatter("t1", src, bins, sizeof(T), management);

// Run histogram on "t1" and produce "t2"
simple_pim_array_red("t1", "t2", sizeof(T), bins, handle, management); 

\end{lstlisting}
\end{figure}

In summary, SimplePIM improves programming productivity for PIM systems. It reduces the number of lines of code required for an application and abstracts away the underlying architectural complexities, such as managing a {scratchpad or} software-managed cache, synchronizing PIM threads/cores, determining data transfer sizes and alignments, and allocating/de-allocating PIM cores. By doing so, SimplePIM makes PIM programming more {programmer}-friendly and accessible to a wider range of developers.

\subsection{Performance {Evaluation}}

\begin{figure}[h]
\centering
\includegraphics[width=\linewidth]{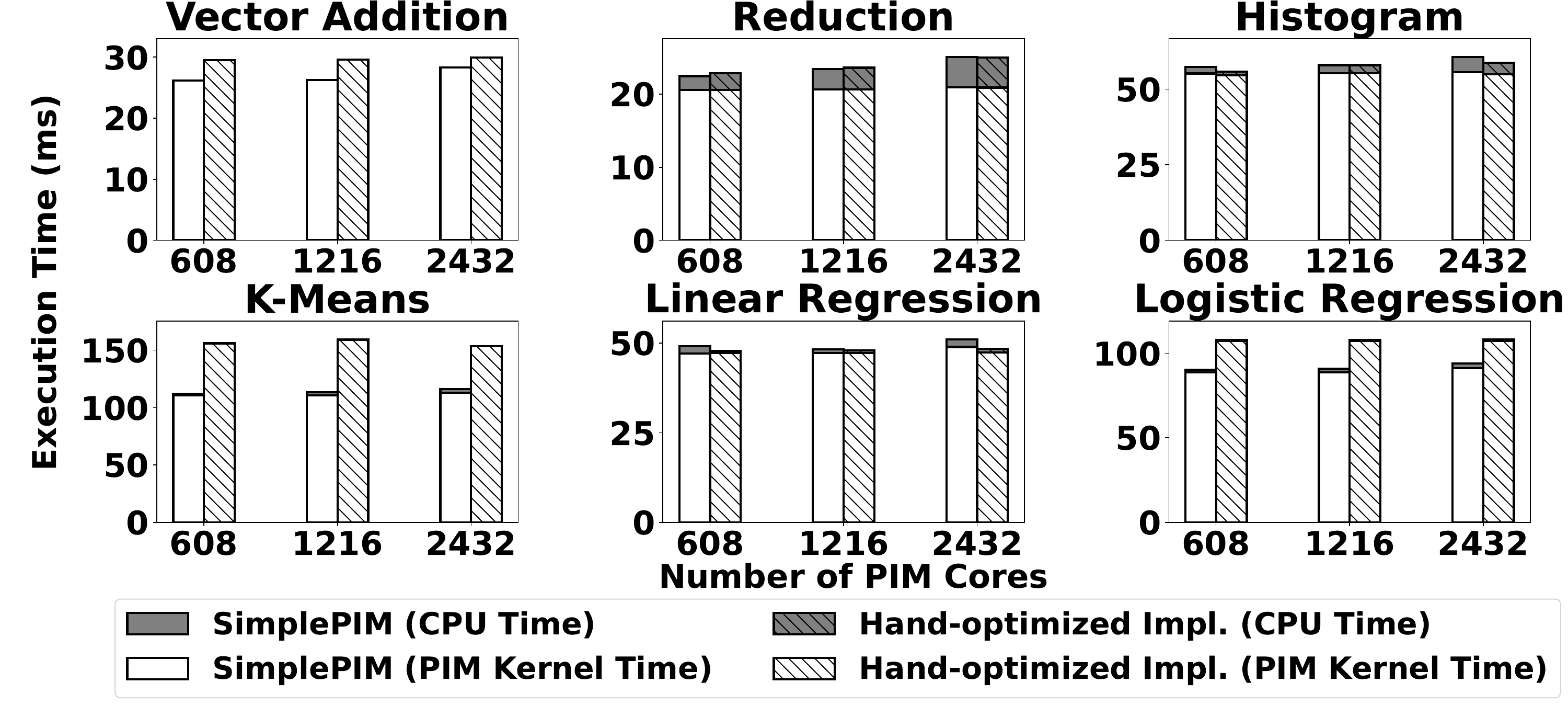}
%\vspace{-28pt}
\caption{Weak Scaling {results for six workloads.}} %SP = SimplePIM implementation, H = Hand-optimized implementation}} 
\label{fig:weak}
%\vspace{-5pt}
\end{figure}

\begin{figure}[h]
\centering
\includegraphics[width=\linewidth]{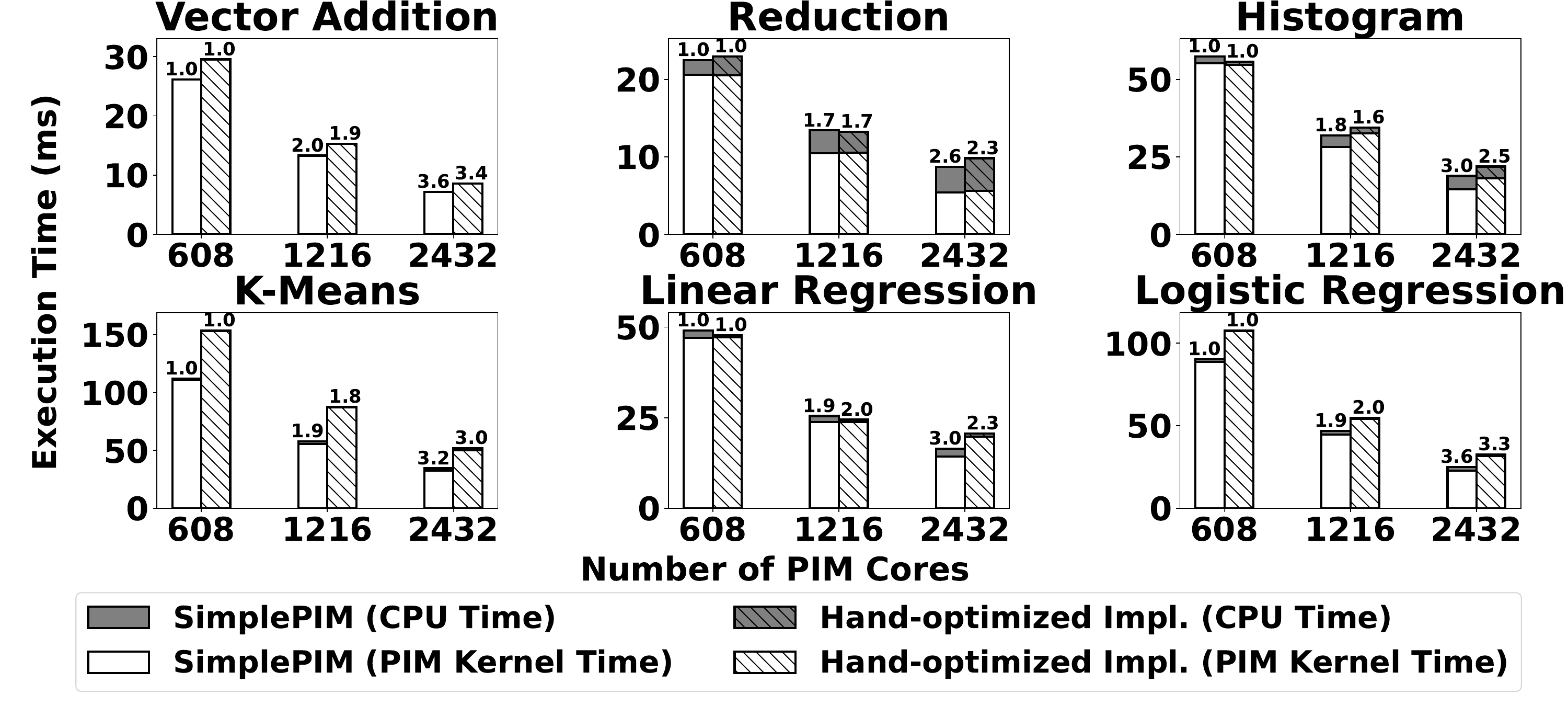}
%\vspace{-18pt}
\caption{Strong Scaling {results for six workloads. The numbers on each bar represent the speedup over 608 PIM cores.}} %SP = SimplePIM implementation, H = Hand-optimized implementation}} 
\label{fig:strong scalability}
%\vspace{-5pt}
\end{figure}

To compare the performance of our SimplePIM code to the baseline code, we {conduct} experiments on {an UPMEM} system with 2,432 PIM cores, and {measure} weak and strong scaling results for each workload on 608, 1,216, and 2,432 PIM cores, as shown in Fig.~\ref{fig:weak} and Fig.~\ref{fig:strong scalability}, respectively. %\jgl{Mention them in order.} \jfc{done} 
The number of elements for each workload is similar to the baseline papers. The number of elements for strong scaling tests is set to be equal to the number of elements \juang{used} for 608 cores in our weak scaling tests. %\jgl{Place figures after the first paragraph that mentions them. Currently both Figs. 6 and 7 are misplaced.} \jfc{done}

%\jgl{These large captions make no sense. The caption should be short. The explanations should be in the text.} \jfc{made smaller}

Overall, the experimental results shown in Fig.~\ref{fig:weak} and Fig.~\ref{fig:strong scalability} demonstrate that SimplePIM achieves comparable performance {to the baseline approach} for \texttt{reduction}, \texttt{histogram}, and \texttt{linear regression}, while consistently outperforming the baseline approach for \texttt{vector addition}, \texttt{logistic regression}, and \texttt{k-means} in both weak and strong scaling experiments. We attribute these performance gains to the various optimization techniques implemented in our framework, as discussed in Section~\ref{sec:st-impl}.

In the weak scaling {evaluation results shown} in Fig.~\ref{fig:weak}, increasing the number of PIM cores linearly with the input size does not significantly impact the performance of both SimplePIM and {hand-optimized} code for all six applications. Furthermore, on average, SimplePIM outperforms the {hand-optimized implementation} in \texttt{vector addition}, \texttt{logistic regression}, and \texttt{k-means} by 1.10$\times$, 1.17$\times$, and 1.37$\times$.

{In the strong scaling evaluation results} shown in Fig.~\ref{fig:strong scalability}, achieving linear speedup with additional PIM cores is not guaranteed, as communication overheads can become dominant. For the \texttt{reduction} workload, which has less kernel execution and comparatively more communication costs, {we observe} only 1.6$\times$ and 2.6$\times$ speedup for 2$\times$ and 4$\times$ more PIM cores. However, for the other five workloads, SimplePIM {achieves} more than 1.8$\times$ speedup with a 2$\times$ increase in PIM cores and 3$\times$ speedup with a 4$\times$ increase in PIM cores.
% \ie{What abount the human code for lin\_reg? It scales poorly at 4$\times$.}
SimplePIM consistently outperforms the hand-optimized {implementations of} all benchmarks, except for \texttt{reduction} with a slight increase in communication cost. SimplePIM outperforms the {hand-optimized implementations} of \texttt{vector addition}, \texttt{logistic regression}, and \texttt{k-means} by 1.15$\times$, 1.22$\times$, and 1.43$\times$ on average {across different numbers of PIM cores}. These results demonstrate the effectiveness of our framework for programming PIM systems in terms of performance.

We {note that, while SimplePIM provides speedup over baseline implementations of} some benchmarks, the performance of {hand-optimized} code can {potentially be equal to or even better than that generated by SimplePIM. 
This requires the programmer to take similar steps as SimplePIM to optimize the code, as Section~\ref{sec:st-impl} describes}. 
Achieving such performance requires careful consideration {and use} of various optimization techniques.
SimplePIM frees \juang{the programmer} from this burden, {thereby} making them more productive.

\subsection{{Evaluation of SimplePIM Variants of Array Reduction}}\label{sec:reduction-eval}

%Recall from Section~\ref{sec:reduction-impl} that SimplePIM selects between two implementations of reduction that use shared and thread-private accumulators.
{SimplePIM provides two variants of \texttt{PIM array reduction} (Section~\ref{sec:reduction-impl}). One variant uses shared accumulators (i.e., one shared output array for all threads and one lock per array entry) and the other one thread-private accumulators (i.e., one output array per thread). {We} compare these two versions {of SimplePIM on the \texttt{histogram} benchmark} with different histogram sizes.

Fig.~\ref{fig:red_comp} shows {the effect of the two versions on the} end-to-end performance of {the \texttt{histogram} benchmark as we vary the number of bins \juang{in the histogram}. 
We make several observations}. 

\begin{figure}[h]
\centering
\includegraphics[width=\linewidth]{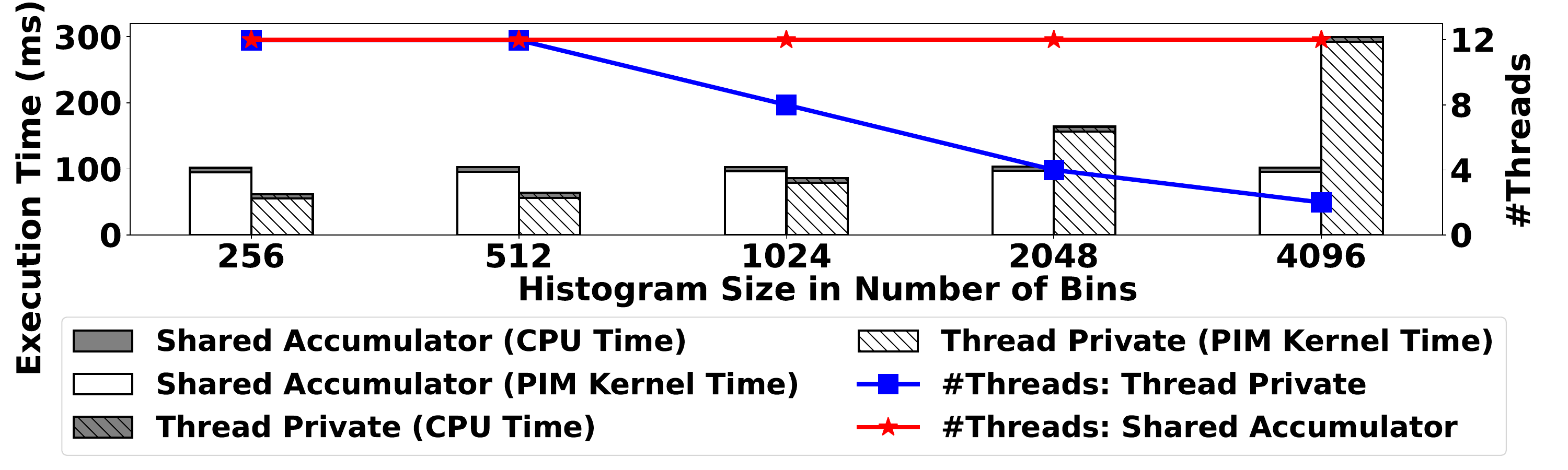}
%\vspace{-12pt}
\caption{Execution time of SimplePIM's shared accumulator version and thread-private version {for the \texttt{histogram} benchmark. Red} and blue lines represent the numbers of active PIM threads.} 
\label{fig:red_comp}
%\vspace{-5pt}
\end{figure} 

First, {the thread-private version is faster than the shared accumulator version for histograms of 256, 512, and 1024 bins. 
Since each thread owns a private output array, there is no need for locks, which avoids the synchronization overhead of the shared memory version. 
When 12 threads are active (for 256-, and 512-bin histograms), the thread-private version outperforms the shared accumulator version by $1.70\times$.} 

Second, the shared accumulator version outperforms the thread-private version for histograms of 2048 and 4096 bins. 
The cause is a reduction in the number of active PIM threads of the thread-private version (blue line) after 1024 bins. 
This reduction relates to the occupancy of the scratchpad memory. For $t$ threads with private histograms of $n$ bins (each of $d$ bytes), $t \times n \times d$ bytes of the scratchpad memory are occupied. 
When the scratchpad size (e.g., 64KB in current UPMEM chips) is not enough for the private histograms (plus buffers for \juang{the} input array), \juang{we should reduce} the number of active threads, as we observe for 1024-, 2048-, and 4096-bin histograms. 
The reduction in the number of active threads causes a linear increase of the execution time, because the pipeline of the PIM cores is not fully busy~\cite{gomez2022benchmarking, prim}. 
As a result, we observe that the execution time of the 2048-bin \texttt{histogram} (with 4 threads) is roughly twice as high as that of the 1024-bin \texttt{histogram} (with 8 threads). \juang{Ditto for} the 4096-bin \texttt{histogram} (with 2 threads) \juang{versus the 2048-bin \texttt{histogram}}.

%% file: secs/6-Discussion.tex
\section{{Discussion}}

While SimplePIM is currently implemented for the UPMEM PIM architecture~\cite{upmem}, it is devised for a broader set of real PIM architectures {(e.g.,~\cite{upmem, axdimm_db, FIMDRAM, lee2021hardware, ke2021near, lee2022isscc, niu2022isscc})}, which have {one or more} \juang{common} characteristics~\cite{gomez2022experimental, item2023transpimlib}: (1) there is a host processor with access to standard main memory and PIM-enabled memory, (2) PIM processing elements (PEs) %are wimpy and 
may need to communicate via the host processor, and (3) the number of PIM PEs scales with memory capacity.

With these characteristics in mind, SimplePIM supports host-PIM communication primitives and its management interface runs on the host. 
SimplePIM's PIM-PIM communication primitives emulate the communication between PIM cores by transparently handling the communication via the host. 
Given that PIM-PIM communications play a pivotal role in simplifying and enhancing PIM programming for high performance, we recommend {the research and development of} more efficient PIM-PIM communication mechanisms in hardware, {e.g.,} as proposed in~\cite{ABC_DIMM, DIMM_Link, rowclone, chang2016low, wang2020figaro, rezaei2020nom}.

SimplePIM can accommodate a wide range of computation patterns beyond the current \texttt{map}, \texttt{reduce}, and \texttt{zip} operations. 
Other parallel patterns, such as prefix sum and filter~\cite{prim, gomez2022benchmarking, gomezluna2021repo}, can be easily incorporated. 
However, extending support to more complex patterns such as stencil and convolution would necessitate a more fine-grained scatter-gather mechanism to handle halo cells {between tiles mapped onto different PIM cores}. %, which may not operate efficiently on the UPMEM PIM architecture. 
Similarly, applications with irregular memory access {patterns}, such as tree data structures~\cite{PimTree}, are also hard to support {due to random access patterns}. 
Future work {can} extend SimplePIM's capabilities by supporting, implementing, and benchmarking additional communication primitives, iterator functions, and workloads on a variety of PIM systems.

%% file: secs/7-RelatedWork.tex
\section{Related Work}

{To our knowledge, SimplePIM is the first {high-level} programming framework specifically designed for real PIM systems. 
{We} first review recent studies of real PIM systems. {We then discuss} several PIM works that propose programming interfaces and compilers for PIM architectures.}

\paragraph{Studies of Real PIM Systems} 
The UPMEM PIM architecture~\cite{upmem} is the first commercially available PIM hardware. 
{Several recent works study this architecture and its suitability to different modern applications. 
Gómez-Luna et al.~\cite{prim, gomez2022benchmarking, upmem_benchmark} present a microbenchmark-based analysis of the UPMEM PIM architecture and a workload suitability study with the PrIM benchmark suite~\cite{gomezluna2021repo}, which contains workloads from dense and sparse linear algebra, machine learning, bioinformatics, image processing, graph processing, etc. 
Nider et al.~\cite{case_study} analyze the UPMEM PIM system for encryption/decryption, compression/decompression, hyper-dimensional computing, and text processing. 
Other works focus on specific applications or application domains, such as 
sparse matrix vector multiplication~\cite{sparsep, giannoula2022sigmetrics}, 
bioinformatics~\cite{DNA_upmem, diab2022hicomb, diab2023framework, lavenier2020, abecassis2023gapim, chen2023uppipe}, 
machine learning~\cite{gomez2022machine, gomez2023evaluating, gomez2022experimental, das2022implementation}, 
transcendental functions~\cite{item2023transpimlib}, 
databases~\cite{lim2023design, PimTree, bernhardt2023pimdb, baumstark2023adaptive, baumstark2023accelerating}, 
homomorphic encryption~\cite{gupta2023homomorphic}, 
and skyline computation~\cite{skyline}.}

There {are also application studies on other real PIM systems~\cite{axdimm_db, FIMDRAM, lee2021hardware, ke2021near, lee2022isscc}. 
Ke et al.~\cite{ke2021near} evaluate sparse embedding operators of \juang{deep-learning-based} recommendation inference~\cite{naumov2019deep} on AxDIMM. 
Lee et al.~\cite{axdimm_db} implement database scan \juang{operations} on AxDIMM. 
Ibrahim and Aga~\cite{ibrahim2023collaborative} implement FFT for commercial PIM architectures, such as Samsung HBM-PIM~\cite{FIMDRAM, lee2021hardware} and SK Hynix AiM~\cite{lee2022isscc} (but the evaluation is done with a performance model).}

\paragraph{Programming Interfaces for PIM}
Several works propose programming interfaces for processing-near-memory architectures (i.e., PIM architectures with processing elements near the memory arrays~\cite{modern_primer}). 
{One approach is \juang{to} use specialized PIM instructions {(e.g., as in~\cite{ahn.pei.isca15, nai2017}) and integrate them into the existing general-purpose sequential execution model}. This approach is {especially} suitable for PIM architectures %with processing elements near the memory banks. %that operate on a single cache line. 
\juang{where communication across PIM processing elements is not possible or easy.} 
When the host processor finds a specialized PIM instruction in the program, it offloads the execution to the PIM processing elements. 
Another approach is to use remote function calls via message passing between different PIM cores {(e.g., as in~\cite{pim_graph})}. This approach is suitable for coarse-grained PIM accelerators with multiple PIM cores that can communicate over an interconnection network.}

{For processing-using-memory architectures (i.e., PIM architectures that compute by leveraging the analog {operational properties} of memory components~\cite{modern_primer}), there are several works that facilitate programming. 
SIMDRAM~\cite{hajinazarsimdram} provides a framework to generate user-defined operations that are executed \juang{via} the simultaneous activation of rows inside DRAM subarrays~\cite{ambit, Seshadri:2015:ANDOR}. 
pLUTo~\cite{ferreira2022pluto} proposes a LUT-based processing-using-DRAM substrate with a compiler that maps complex operations onto LUT queries.}

\paragraph{Compilers for PIM}
{Several works propose compilers for simulated PIM architectures. 
Duality Cache~\cite{fujiki2019duality} proposes a compiler that accepts existing CUDA programs and maps the computation onto a processing-using-SRAM substrate. 
Infinity Stream~\cite{wang2023infinity} proposes an intermediate representation and a just-in-time compiler for processing-near-memory, processing-using-memory, and host execution. 
\juang{CHOPPER~\cite{peng2023chopper} presents a bit-serial compiler for processing-using-DRAM substrates.} 
CINM~\cite{khan2022cinm} is a compiler flow for simulated processing-using-memory architectures and processing-near-memory architectures such as UPMEM. It is based on MLIR~\cite{lattner2020mlir, lattner2021mlir} and it supports linear algebra PIM kernels. 
SimplePIM is the first programming framework for real PIM architectures that supports a {wide} variety of PIM kernels.}

%% file: secs/8-Conclusion.tex
\section{Conclusion}

{We introduce SimplePIM, which is a new high-level programming framework for real PIM systems. 
SimplePIM is specifically \juang{designed} to enable efficient and productive programming.} 
%SimplePIM is the first programming framework specifically designed for real PIM architectures. 
SimplePIM {efficiently} leverages the host CPU for data management, and incorporates primitives for PIM-host communication (e.g, \texttt{scatter}, \texttt{gather}, \texttt{broadcast}) and primitives for {communication among} PIM cores (e.g., \texttt{allreduce}, \texttt{allgather}). 
SimplePIM's easy-to-use interface provides iterators (e.g., \texttt{map}, \texttt{reduce}, \texttt{zip}) that allow programmers to largely avoid the complexity of the underlying PIM architecture. 
Our implementation of SimplePIM for the UPMEM system enables programmers to develop PIM programs with {$2.98\times$ to $5.93\times$} less code than hand-optimized programs, while providing equal or higher performance. %than them. 
SimplePIM frees programmers from {dealing with the complexities and idiosyncracies} of \juang{the low-level PIM} hardware, while {also enabling programmer-transparent \juang{code} optimizations}. 
We believe that SimplePIM is a milestone to ease the programmability {and adoption} of current and future real PIM systems. %, and we hope to see more work in this direction. 
{We hope that more future work ensues in programming memory-centric systems, and to facilitate that we open-source our SimplePIM framework at \url{https://github.com/CMU-SAFARI/SimplePIM}}.